\documentclass[twocolumn,preprintnumbers,amsmath,amssymb]{revtex4}

\usepackage{graphicx}
\usepackage{dcolumn}
\usepackage{bm}
\usepackage[dvips]{color}

\newcommand{\kB}{k_{\mathrm{B}}}
\newcommand{\kT}{k_{\mathrm{T}}}

\newcommand{\Np}{N_{\mathrm{p}}}

\renewcommand{\vec}[1]{\ensuremath{\bm{#1}}}

\newcommand{\ULJ}{{\ensuremath{U_{\mathrm{LJ}}}}}

\newcommand{\UFENE}{{\ensuremath{U_{\mathrm{FENE}}}}}

\newcommand{\Rg}{{R_{\text{g}}}}

\newcommand{\Rgbulk}{R_{\rm g}^{\rm bulk}}

\newcommand{\Rgpar}{R^{2}_{\rm g,\parallel}}
\newcommand{\Rgper}{R^{2}_{\rm g,\bot}}

\newcommand{\Reepar}{R^{2}_{\rm ee,\parallel}}
\newcommand{\Reeper}{R^{2}_{\rm ee,\bot}}

\newcommand{\diameter}{d}

\newcommand{\rc}{\ensuremath{r_{\mathrm{c}}}}

\newcommand{\tauLJ}{\tau_{\mathrm{LJ}}}

\newcommand{\Tc}{\ensuremath{T_{\mathrm{c}}}}

\newcommand{\Tg}{\ensuremath{T_{\mathrm{g}}}}

\newcommand{\rbond}{\ensuremath{r_{\mathrm{b}}}}

\newcommand{\rmin}{\ensuremath{r_{\mathrm{min}}}}

\newcommand{\fref}[1]{Fig.\ \ref{#1}}

\newcommand{\eref}[1]{Eq.\ (\ref{#1})}

\renewcommand{\vec}[1]{\bm{#1}}

\newcommand{\bfig}{\begin{figure}}
\newcommand{\efig}{\end{figure}}

\newcommand{\rsfig}[2]{
\begin{center}
\includegraphics*[width=#1\textwidth]{#2}
\end{center}
}

\newcommand{\remove}[1]{}

\begin{document}

\preprint{Invited feature article to appear in IJMR (2009)}
\title{Multiscale modeling of polymers at interfaces}

\author{Fathollah Varnik$^1$, Kurt Binder$^2$}
\address{
$^1$Interdisciplinary Center for Advanced Materials Simulation (ICAMS), Ruhr University Bochum,
Stiepeler Stra{\ss}e 129, 44801 Bochum, Germany
\\
$^2$Institut f\"ur Physik, Johannes Gutenberg-Universit\"at Mainz, Staudinger Weg 7, 55099 Mainz, Germany}

\date{\today}

\begin{abstract}
A brief review of modeling and simulation methods for a study of polymers at interfaces is provided.
When studying truly multiscale problems as provided by realistic polymer systems, coarse graining is practically unavoidable. In this process, degrees of freedom on smaller scales are eliminated to the favor of a model suitable for efficient study of the system behavior on larger length and time scales. We emphasize the need to distinguish between dynamic and static properties regarding the model validation. A model which accurately reproduces static properties may fail completely when it comes to the dynamic behavior of the system. Furthermore, we comment on the use of Monte Carlo method in polymer science as compared to molecular dynamics simulations. Using the latter approach, we also discuss results of recent computer simulations on the properties of polymers close to solid substrates. This includes both generic features (as also observed in the case of simpler molecular models) as well as polymer specific properties. The predictive power of computer simulations is highlighted by providing experimental evidence for these observations. Some important implications of these results for an understanding of mechanical properties of thin polymer films and coatings are also worked out.
\end{abstract}
\pacs{64.70.Pf, 05.70.Ln, 83.60.Df, 83.60.Fg}
\maketitle
\section{Introduction}
Due to their ubiquitous presence in everyday life, the importance of polymers can hardly be over estimated. Indeed, polymer science has had a major impact on the way we live. Just 50 years ago, materials we now take for granted did not exist at all. Due to their structural complexity, polymers are generally not crystalline at low temperatures. Rather, they exhibit an amorphous, glassy structure. The glass transition (the transition from a liquid to an amorphous solid) is thus an important concept when it comes to an understanding of the properties of polymer systems.

As to the application, polymers are often used as protective coatings in many fields ranging from the car industry (corrosion resistance) to microelectronics (protection against thermal as well as mechanical load)~\cite{Zhang::PolEngSci::1999, Rayss::JApplPolSci::1993, Armstrong::ElectrochemicaActa::1993}. 

In such situations, the polymer is always in contact with a solid substrate or even confined between two solid plates (film geometry). An important piece of information for materials design is therefore how the thermal and mechanical properties of a polymer system are affected by interaction with a solid substrate.

In addition to its technological importance, the investigation of polymers close to solid surfaces is also of great theoretical interest. One is, for example interested to understand how the broken symmetry of the space in the proximity of the substrate as well as related energetic issues modify static and dynamic properties (chain conformation, diffusion, reorientation dynamics, etc.) of polymers close to the interface. A particularly interesting aspect is also how the glass transition is affected by the presence of the solid substrates.

However, the presence of a wide range of time and length scales makes the study of polymer systems a very challenging multiscale problem \cite{LodgeMuthu_JPC1996}.  In the simplest case of linear homopolymers, each macromolecule contains $\Np$ identical repeat units (monomers), connected to form a chain. In experiments, the chain length may vary in the range $10^2 \lesssim \Np \lesssim 10^6$.  This implies that the average size of a polymer, measured for instance by the radius of gyration $\Rg$ \cite{DoiEdwards,RubinsteinColby},  varies between $\Rg \sim\! 10 d$ up to $\Rg \sim\! 1000 d$, where $d$ denotes the diameter of a single monomer ($d$ being of the order of a few $\AA$).

These different length scales are reflected in the particular
features of a polymer melt.  In the melt the monomers pack densely,
leading to an amorphous short-range order on a local scale and to an
overall low compressibility of the melt. Both features are
characteristic of the liquid state.  Qualitatively, the collective
structure of the melt thus agrees with that of non-polymeric
liquids. Additional features, however, occur if one considers the
scale of a chain.  A long polymer in a (three-dimensional) melt is
not a compact, but a self-similar object
\cite{RubinsteinColby,DeGennes,Witten_ROP1998}.  It possesses a
fractal `open' structure which allows other chains to penetrate into
the volume defined by its radius of gyration.  On average, a polymer
interacts with $\sqrt{\Np}$ other chains,
a huge number in the large-$\Np$ limit.  This strong
interpenetration of the chains has important consequences.  For
instance, intrachain excluded volume interactions, which would swell
the polymer in dilute solution, are screened by neighboring chains
\cite{DoiEdwards,RubinsteinColby,DeGennes,Edwards_JPhysA1975,MuthukumarEdwards_JCP1982,WittmerEtal:PRL2004}, although nontrivial nongaussian orientational correlations due to excluded
volume remain \cite{WittmerEtal:PRL2004}.

A polymer in a melt thus in many respects behaves on large scales as if it were a
random coil, implying that its radius of gyration scales with chain
length like $\Rg \sim \sqrt{\Np}$.  Furthermore, the
interpenetration of the chains creates a temporary network of
topological constraints
\cite{DoiEdwards,RubinsteinColby,DeGennes,McLeish_AdvPhys2002}.
These entanglements greatly slow down the chain dynamics and render
the melt in general very viscous compared to low-molecular weight liquids.

In this article, we will provide a brief overview on computer simulations of polymers close to interfaces, while at the same time briefly touching upon some of the techniques aimed at a reduction in the computational cost related to the presence of multiple length and time scales. It should, however, be emphasized here that the present survey is far from being exhaustive, neither in regard to the modeling of polymer melts nor in regard to specific computational aspects. We refer the interested reader to extended reviews on these topics, e.g.\ references~\cite{MullerPlathe:2002,MullerPlathe:2003,GlotzerPaul:AnnRevMatSci2002,Kroger:PhysRep2004,Binder_MCMD1995,BinderBaschnagelPaul2003,BaVa2005}.

In principle, a full multiscale simulation approach constructs the connection between the quantum description of interactions (Fig.~1) over the all-atom description of a chemically realistic model but using effective classical potentials up to the coarse-grained level of mesoscale models explicitly. Such an explicit multiscale approach which derives effective interactions on the mesoscale ab initio still is an ambitious and difficult task 
\cite{DelleSite2005a,DelleSite2005b,Praprotnik2008,Mulder2009}
and most of the work in the literature considers only partial steps of the full problem. Thus, in this article, we will concentrate on the description in terms of coarse-grained bead-spring type models lacking explicit chemical details, and we shall describe the steps (Fig.~1) to derive these models in a schematic way only.

\section{Coarse-graining}
\label{subsec:cg} On a fundamental level, interaction forces originate from the adaption of the electronic degrees of freedom to the positions of the nuclei.  It may therefore appear natural to model polymer melts via the Car--Parrinello method \cite{MarxHutter_NIC2000}.  This method is a molecular dynamics (MD) technique \cite{Allen_NIC2004,BinderEtal_JPCM2004} which allows the electrons to adiabatically follow the motion of the nuclei, thereby replicating the energy landscape that the nuclei feel at any instant of their motion. Using recent extensions of the method, one is now able to  study system sizes up to about 1000 nuclei for about $10$ ns \cite{Kuehne_PRL2007}. This time, however, barely suffices to equilibrate the system at high temperature in the liquid state \cite{Binder_MCMD1995,Kremer:MacroChemPhys2003,BaVa2005}.

\begin{figure} \rsfig{0.45}{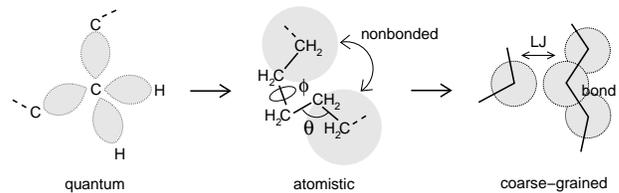} 
\caption[]{A schematic view of different scales which can be focused upon when describing polymers. The quantum level takes account of the electrons to calculate the interactions between the nuclei. On the computationally less demanding atomistic level, the electronic degrees of freedom are replaced by a force field  \cite{PaulSmith_RPP2004}.  Computationally still less demanding than atomistic models are simulations at the coarse-grained level. Here, a monomer is associated with a spherical site and the realistic potentials are replaced by simpler ones~\cite{JBETal:AdvPolySci2000,MullerPlathe:2002,MullerPlathe:2003,BaVa2005}.} 
\label{fig:coarse_graining} \end{figure}

Some kind of coarse-graining procedure is therefore necessary in order to adequately describe statistical mechanical properties of polymer systems on time and length scales of interest. Such a procedure usually consists of elimination of fast degrees of freedom by incorporating them in effective potentials \cite{JBETal:AdvPolySci2000,MullerPlathe:2003,PaulSmith_RPP2004}.

\subsection{Atomistic models}
A first degree of simplification may consist in replacing the electronic degrees of freedom by empirical potentials for the bond lengths, the bond angles, the torsional angles and the nonbonded interactions between distant monomers along the chain (`quantum level $\rightarrow$ atomistic level', see \fref{fig:coarse_graining}). This step introduces a `force field', i.e., the form of the potentials is postulated and the corresponding parameters (e.g.\ equilibrium bond length, force constants, etc.) are determined from quantum-chemical calculations and experiments \cite{JBETal:AdvPolySci2000,PaulSmith_RPP2004}.

Several such force fields have been proposed throughout the past decades for both explicit atom models and united atom models.  An explicit atom model treats every atom as a separate interaction site, whereas a united atom model lumps a small number of real atoms together into one site \cite{JBETal:AdvPolySci2000,MullerPlathe:2003,PaulSmith_RPP2004}. Typical united atoms are CH, $\mathrm{CH}_2$, and $\mathrm{CH}_3$. The reduction of force centers translates into the computational advantage of allowing longer simulation times.  With a time step of $\sim\! 10^{-15}$ s---compared to $\sim\! 10^{-17}$ s for the Car-Parrinello method---a few thousand united atoms can be simulated over a time lapse of several $\mu$s, about an order of magnitude longer than an explicit atom simulation of comparable system size.

Both explicit atom models and united atom models have been used in the study of glass-forming polymers (see e.g.\ \cite{Clarke_review1995,Clarke:currentopinion1998} for reviews on older work).  Current examples include polyisoprene (explicit atom; \cite{ColmeneroEtal:PRE2002,ColmeneroEtal:JPCM2003}), atactic polystyrene (united atom; \cite{LyulinEtal:Macromolecules2002_1,LyulinEtal:Macromolecules2002_2,LyulinEtal:Macromolecules2003}) and {\em cis-trans} 1,4-polybutadiene (united and explicit atom models; \cite{PaulSmith_RPP2004,SmithEtal:JCP2004,KrushevPaul_PRE2003,KrushevEtal_Macromolecules2002,SmithEtal:JCP2002,GrantPaul_ChemPhys2000}). Certainly, the ultimate objective of these modeling efforts is that the simulation results lend themselves to a quantitative comparison with experiments.  Such a comparison may, however, require a careful fine-tuning of the force field.  For the family of neutral hydrocarbon polymers the optimization of the torsional potential appears particularly crucial.  Not only the position and the relative depth of the minima, but also the barriers between them should be accurately determined, as local relaxation processes, involving transitions between the minima, are exponentially sensitive to them. In extreme cases, imprecise barrier heights may seriously affect the dynamics while leaving structural features of the melt unaltered.

\begin{figure}
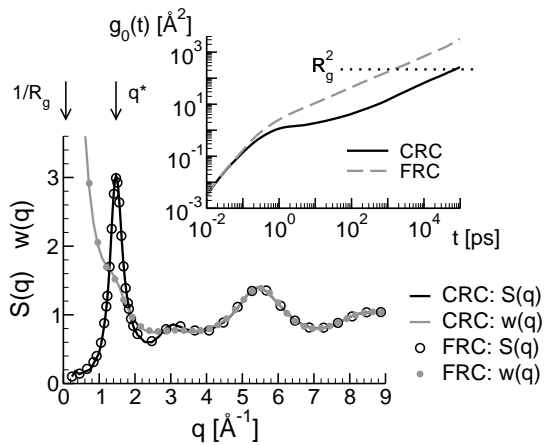
 \rsfig{0.4}{PB_SqWq_MSD.eps} \caption[]{Simulation results for
{\em cis-trans} 1,4-polybutadiene (adapted from reference~\cite{KrushevPaul_PRE2003}).  Main panel: Collective static structure factor, $S(q)$, and single-chain  structure factor, $w(q)$, versus the modulus of the wave vector $\vec{q}$ at $T = 273$ K. Two united atom models are compared: a chemically realistic model (CRC) and the same model but without torsional potential (FRC).  The vertical arrows indicate the $q$-values associated with the radius of gyration and with the first maximum of $S(q)$ (`amorphous halo'). The maximum occurs at $q^*\simeq 1.47$ {\AA}$^{-1}$.  In real space, this value would correspond to an intermonomer distance of $\approx 4.3$ {\AA} which is roughly compatible with the average Lennard--Jones diameter of the model ($\diameter \approx 3.8$ {\AA}). Inset: Mean-square displacement $g_0(t)$, averaged over all monomers, versus time for the CRC and FRC models at $T= 273$ K. The horizontal line indicates the radius of gyration $\Rg^2= 218$ {\AA}$^2$ (which is found to be the same for both models \cite{KrushevEtal_Macromolecules2002}).}
\label{fig:PBD_CrcFrc}
\end{figure}

Such an example is shown in \fref{fig:PBD_CrcFrc}.  The figure compares simulation results for two models of a polybutadiene melt:  a carefully validated united atom model which reproduces the experimentally found structure and dynamics of the melt, and the same model with the torsional potential switched off.  Apparently, suppression of the torsional potential has no influence on the structure, but considerably accelerates the monomer dynamics.

The above example demonstrates that different potentials may lead to realistic representations of structural properties, but to diverging predictions for the dynamics.  Such an observation is not limited to polymers, but was also made e.g.\ for amorphous $\mathrm{SiO}_2$ \cite{kobreview1999}.  It suggests that the design of a chemically realistic model, aiming at a parameter-free comparison between simulation and experiment, should involve information about both the structural and the dynamic properties.

\subsection{Generic models}
In many cases, one is interested in a basic understanding of polymeric features rather than detailed description of a specific model. For uncharged linear polymers, these features are presumed to be the dependence of material properties on chain connectivity, excluded-volume interactions, monomer--monomer attractions and/or some stiffness along the chain backbone.  In this case, it appears permissible to forgo fast degrees of freedom (bond length and bond angle vibrations, etc.) in favor of a coarse-grained model. A special type of coarse-grained models are so-called `generic models' \cite{MullerPlathe:2003}.  Various such generic models have been studied in the literature (for reviews see references~\cite{Binder_MCMD1995,KremerBinder1988,BaschnagelWittmerMeyer:NIC_Review2004,Kotelyanskii2004}). Due to the lack of space, in the following we present only one of these models in more detail, which was used in recent simulations \cite{BennemannPaulBinder1998,BennemannBaschnagelPaul1999_incoherent,BennemannPaulBaschnagel1999,BennemannPaulBaschnagel1999_Rouse,natureBDBG1999,betaDynamics,alphaDynamics,BuchholzPaulVarnikBinder:JCP2002,AicheleEtal_2002,AicheleEtal_PRE2004,VarnikEtal:JCP2000,Varnik:CPC2002,VarnikEtal:PRE2002,VarnikEtal:EPJE2002,VarnikBinder:JCP2002,VarnikEtal:EPJE2003,Peter2006,Peter2007,Peter2008}. Certainly, this choice is biased by our own experience.

Indeed, the issue of polymers at interfaces in particular and confined liquids in general has received considerable attention in simulation studies. Various systems---simple liquids \cite{ScheidlerEtal:EPL2000,ScheidlerEtal:EPL2002,ScheidlerEtal:JPCB2004,GalloEtal:EPL2002,FehrLoewen:PRE1995,BoddekerTeichler:PRE1999}, hydrogen-bonded or molecular liquids \cite{GalloEtal:JCP2000,TeboulSimionesco:JPCM2002}, silica \cite{Roder:JCP2001}, polymers \cite{XuMattice:Macro2003,StarrEtal:PRE2001,StarrEtal:Macro2002,YoshimotoEtal:JCP2005,JaindePablo:PRL2004,JaindePablo:Macro2002,BohmedePablo:JCP2002,torresEtal:PRL2000,MansfieldEtal:Macro1991,ManiasEtal:ColSurf2001,BaljonEtal:Macro2005,BaljonEtal:PRL2004}---and confining geometries---pores \cite{ScheidlerEtal:EPL2000,GalloEtal:JCP2000,TeboulSimionesco:JPCM2002}, fillers in glass-forming matrices \cite{StarrEtal:PRE2001,StarrEtal:Macro2002,GalloEtal:EPL2002}, thin films \cite{XuMattice:Macro2003,YoshimotoEtal:JCP2005,JaindePablo:PRL2004,JaindePablo:Macro2002,BohmedePablo:JCP2002,torresEtal:PRL2000,MansfieldEtal:Macro1991,ManiasEtal:ColSurf2001,BaljonEtal:Macro2005,BaljonEtal:PRL2004,Roder:JCP2001,ScheidlerEtal:EPL2002,ScheidlerEtal:JPCB2004,FehrLoewen:PRE1995,BoddekerTeichler:PRE1999}---have been considered. References provided here as well as throughout this paper are hoped to, at least partially, compensate this shortcoming.

\section{A bead-spring model for polymer melts}
\label{subsec:beadspring} In 1990 \cite{KremerGrest1990} Kremer and Grest proposed a versatile bead-spring model for the simulation of polymer systems.  The `Kremer--Grest' model has ever since been deployed to investigate numerous problems in polymer physics, including relaxation processes in polymer solutions \cite{DuenwegEtal:KB_review1995} and melts \cite{KremerGrest_review1995,PuetzKremerGrest2000,Kremer:NIC2004} or the behavior of polymer brushes \cite{Grest_review1995,Grest_review1999}, to name just a few.

In a variant of this model, proposed by Bennemann and coworkers \cite{BennemannPaulBinder1998}, the chains contain $\Np$ identical monomers of mass $m$.  All monomers, bonded and nonbonded ones, interact by a truncated Lennard--Jones (LJ) potential
\begin{equation}
\ULJ (r) = \left \{
\begin{array}{ll}
4 \epsilon \left[ (\diameter/r)^{12} - (\diameter/r)^{6} \right] & \quad \mbox{for} \; r \leq \rc\;, \\
0 & \quad \mbox{else} \;.
\end{array}
\right . \label{eq:LJ12-6TS}
\end{equation}
The parameter $\rc = 2\rmin$ is the cut-off distance, where $\rmin = 2^{1/6}\diameter$ is the minimum of \eref{eq:LJ12-6TS}. In contrast to the original version of the Kremer--Grest model,
where $\rc = \rmin$ (leading to purely repulsive intermolecular interactions), the cutoff distance proposed by Bennemann is motivated by the wish to work with a potential that is as short-ranged as possible while yet including the major part of the attractive van-der-Waals interaction. Even though attractive interactions are not expected to appreciably affect the local structure in a dense melt, they have a significant effect on thermodynamic properties.  Furthermore, they are important for simulations of e.g.\ the phase behavior of polymer solutions \cite{MuellerMacDowell:Macro2000,VirnauEtal:NewJP2004}, thin films with a film-air interface \cite{HeineEtal:PRE2003,TsigeGrest:Macromolecules2004} or crazing in polymer glasses \cite{BaljonRobbins:Macro2001,RottlerRobins:PRE2003}.

The chain connectivity is ensured by  a FENE (finitely extensible non-linear elastic) potential
\begin{equation}
\UFENE (r) = - \frac{1}{2}k R_0^2 \ln \Big [1 -
\Big(\frac{r}{R_0}\Big)^2 \Big], \; R_0=1.5\diameter,\,
k=\frac{30\epsilon}{\diameter^2} \; \label{eq:FENE}
\end{equation}
The FENE potential diverges logarithmically in the limit of $r \rightarrow R_0$ (`finite extensibility') and vanishes parabolically as $r \rightarrow 0$ (`elastic behavior'). 
The superposition of the FENE- and the LJ-potentials yields a steep effective bond potential with a minimum at $\rbond \approx 0.96d$ (see for instance Ref.~\cite{VarnikEtal:EPJE2002}).
The difference between $\rbond$ and $\rmin$ is crucial for the ability of the model
to bypass crystallization and exhibit glass-like freezing in.

\subsection{Approximate mapping to real units} The parameters of \eref{eq:LJ12-6TS} define the characteristic scales of the melt:  $\epsilon$ the energy scale, $\diameter$ the length scale, and $\tauLJ = (m \diameter^2 / \epsilon)^{1/2}$ the time scale.  In the following, we utilize LJ-units.  That is, $\epsilon = 1$, $\diameter= 1$, and $m = 1$.   Furthermore, temperature is measured in units of $\epsilon/\kB$ with the Boltzmann constant $\kB = 1$.

Although reduced units are commonly employed in simulations and are
of technical advantage \cite{AllenTildesley,FrenkelSmit}, it might
still be interesting to obtain a feeling how they translate into
physical units.  Such a mapping to real
systems has recently been carried out by Virnau \cite{VirnauEtal:NewJP2004,VirnauEtal:JCP2004} and by Paul and Smith
\cite{PaulSmith_RPP2004}.  Virnau explored the phase
separation kinetics of a mixture of hexadecane
($\mathrm{C}_{16}\mathrm{H}_{34}$) and carbon dioxide
($\mathrm{CO}_2$).  By identifying the critical point of the
liquid-gas transition in hexadecane with that of bead-spring chains
containing 5 monomers they found $\diameter \simeq 4.5 \times 10^{-10}$
m and $\epsilon \simeq 5.8 \times 10^{-21}$ J.  Paul and Smith
used the data on the dynamics of chemically realistic models for nonentangled melts 
of polyethylene and polybutadiene and obtained $\tau_\mathrm{LJ}\simeq 0.21 \times 10^{-12}$s.  
These values for $\diameter$, $\epsilon$, and $\tau_\mathrm{LJ}$ are compatible with the
estimates obtained by Kremer and Grest when comparing the dynamics
of entangled bead-spring melts to real polymers (see table III of
\cite{KremerGrest1990}).\label{page:estimate-for-d-LJ}

\subsection{Choice of the chain length}  In polymer glass simulations the chain length is usually chosen as a compromise between two opposing wishes:  On the one hand, it should be sufficiently large to separate the scales of the monomer and of the chain size so that polymer-specific effects (or at least the onset thereof) become observable.  On the other hand, computational expedience suggests to work with short chains.  Because the simulations aim at following the increase of the monomeric relaxation time $\tau_0$ with decreasing temperature over as many decades as possible, slow relaxation processes, already present at high temperatures ($T$) due to entanglements, should be avoided.  Thus, the chain length should be smaller (or at least not much larger) than the  entanglement length $N_\mathrm{e}$.  Extensive studies of the Kremer--Grest model show that $N_\mathrm{e} \approx 32$ Shorter chains exhibit Rouse-like dynamics \begin{equation} \tau(\Np) = \tau_0 \Np^{\approx 2} \;. \label{eq:tau_Rouse} \end{equation} As the Bennemann model is expected to have a similar $N_\mathrm{e}$, the chain length $\Np = 10$ was proposed as a possible compromise \cite{BennemannPaulBinder1998}.  This chain length was used in all subsequent studies pertaining to glass-forming polymer melts \cite{BennemannPaulBinder1998,BennemannBaschnagelPaul1999_incoherent,BennemannPaulBaschnagel1999,BennemannPaulBaschnagel1999_Rouse,natureBDBG1999,betaDynamics,alphaDynamics,BuchholzPaulVarnikBinder:JCP2002,AicheleEtal_2002,AicheleEtal_PRE2004,VarnikEtal:JCP2000,Varnik:CPC2002,VarnikEtal:PRE2002,VarnikEtal:EPJE2002,VarnikBinder:JCP2002,VarnikEtal:EPJE2003}.

\subsection{Including solid substrates}
\label{subsec:films} 
Even though real substrates can have a complex structure, it appears natural in the spirit of the polymer models discussed above also to treat the substrate at a generic level. One obvious feature is its impenetrability. So a minimal model must at least respect monomer-substrate excluded volume interactions.  Further generic features could be some surface roughness and adhesive power.  Based on this reasoning, simulations often model the substrate as a crystal  \cite{PatrykiejewSurfSci2000,Steele:SurfSci1973} made of particles that interact with each other and with the monomers by LJ-potentials.  

\begin{figure}
\unitlength=1mm
\begin{picture}(100,20)(5,0)
\put(18,-11){
\includegraphics*[width=60mm,clip=]{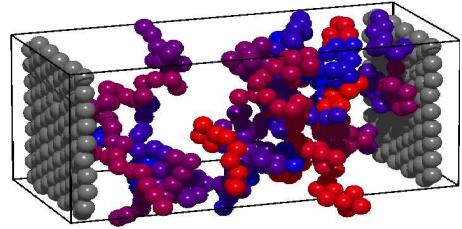}}
\end{picture}
\vspace*{5mm} \caption[]{Snapshot of a polymer system between two substrates of triangular lattice structure (only $40$ chains out of $200$, each containing $\Np = 10$ monomers, are shown).}
\label{fig:polymerfilm-snapshot}
\end{figure}

Such crystalline substrates may be implemented by tethering the substrate ('wall') atoms to the sites of a triangular lattice \cite{VarnikBinder:JCP2002} via harmonic springs (`Tomlinson model' \cite{RobbinsMueser2001,He-Robbins})
\begin{equation}
U_{\mathrm{T}}(\vec{r}) = \frac{1}{2} \kT \big ( \vec{r} -
\vec{r}_{\mathrm{eq}} \big )^2, \quad  \kT = 100 \mathrm{~(LJ~units).}
\label{eq:Tomlinson}
\end{equation}

Here, $\vec{r}_\mathrm{eq}$ denotes the equilibrium position of an atom on the triangular lattice and $\kT$ the spring constant. The substrate atoms are LJ-particles that interact with each other and with the monomers.  The parameters ($\epsilon$ and $\diameter$) for these interactions---wall-wall and monomer-wall---are the same as in \eref{eq:LJ12-6TS}.

Since the spatial arrangement of the substrate atoms may have a strong influence on the properties of the melt in the very vicinity of the substrate, it is interesting to also study the effect of substrates with an amorphous structure. This can be achieved in a way quite similar to the implementation of crystalline walls (Fig.~\ref{fig:polymerfilm-snapshot}), the only difference being the random (instead of regular) distribution of substrate atoms.

If one is only interested in the average force which the substrate exerts on a monomer, one may treat the wall as a continuum and integrate over the parallel $(x,y)$-directions and the vertical one up to the wall-melt interface. Carrying out this calculation for the LJ-potential one obtains 
\begin{equation}
U_{\mathrm{w}}(z) = \epsilon_{\mathrm{w}} \bigg [\Big (\frac{\diameter}{z}
\Big)^{9} - f_{\mathrm{w}} (\frac{\diameter}{z} \Big)^{3} \bigg ]
\label{eq:LJ9-3}
\end{equation}
where $\epsilon_\mathrm{w}$ denotes the monomer-wall interaction energy and $f_{\mathrm{w}}$ is a constant.  While the second attractive term is important if one wants to study polymer adsorption \cite{MetzgerEtal:MacroTheo2002} or wetting phenomena \cite{MetzgerEtal:JCP2003,MuellerEtal:IJMPC2001}, the first term of \eref{eq:LJ9-3} suffices to impose a geometric confinement.  This is the stance we have adopted in most of the simulations on supercooled polymer films \cite{VarnikEtal:JCP2000,Varnik:CPC2002,VarnikEtal:PRE2002,VarnikEtal:EPJE2002,VarnikEtal:EPJE2003}.

\section{Molecular dynamics versus Monte Carlo}
In  the framework of computer simulations, it appears natural to address dynamical problems via MD techniques.  However, if one is interested in equilibrating long-chain glass-forming polymer melts at low $T$, MD might not be the most efficient approach. The realistic molecular dynamics has the drawback that the available simulation time is often not sufficiently long for an equilibration of the polymer configuration at low temperatures or for long chain lengths.

Therefore, one might envisage resorting to Monte Carlo (MC)
techniques \cite{LandauBinder,Binder2008,Binder2009,Binder2009b}.  
The strategic advantage offered by this method is the number of ways in which MC moves may be designed
to explore configuration space.  The hope is to find an algorithm
that, freed of the need to capture the real dynamics, efficiently
decorrelates the configurations of glass-forming polymer melts at
low $T$.  This demand on the algorithm appears to exclude the
simplest MC technique, the application of only local MC moves, as a
possible candidate.  A local MC move consists of selecting a
monomer at random and in attempting to displace it by a small amount
in a randomly chosen direction
\cite{BaschnagelWittmerMeyer:NIC_Review2004}.  Not only should the
local character of coordinate updating share the essential
problematic features of the (local) molecular dynamics at low $T$ or for
large $\Np$, but also may one expect that local MC moves will yield
an unfavorably large prefactor of the relaxation time due to their
stochastic character.  This conjecture is based on an observation
made by Gleim and coworkers \cite{GleimKobBinder1998}. They compared
the relaxation dynamics of a glass-forming binary mixture simulated,
on the one hand, by MD and, on the other hand, by a stochastic
(Brownian) dynamics (which is in some respect similar to MC).  They
demonstrated that, although the structural relaxation at
long times is the same for both methods, MD is roughly an order of
magnitude faster than the stochastic dynamics.

However, MC moves need not be local.  They can be tailored to alter
large portions of a chain.  A prominent example of such nonlocal
moves is the configuration-bias Monte Carlo (CBMC) technique
\cite{FrenkelSmit,BaschnagelWittmerMeyer:NIC_Review2004}.
Application of this technique to dense polymer systems in the
canonical ensemble usually involves the attempt to remove a portion
of a chain starting from one of its monomers that is randomly chosen
and to regrow the removed portion subject to the constraints imposed
by the local potential energy.  If successful, this implies a large
modification of the chain configuration, thereby promising efficient
equilibration.  However, Bennemann  found that even in the
limit where only the end is reconstructed (`smart reptation'), CBMC
is inferior to ordinary MD \cite{BennemannPaulBinder1998}.  In a
dense melt, the probability of inserting a monomer becomes vanishingly
small anywhere except at the position where it was removed.  So, the
old configuration of the chain is just restored.  This trapping of
the chain makes the relaxation become very slow.

Thus, successful nonlocal chain updates in dense systems should
involve moves that do not require (much) empty space.  A promising
candidate is double-bridging algorithms which were successfully
employed in simulations of polyethylene chains
\cite{KarayiannisEtal:JCP2002,BanaszakEtal:JCP2003}, of the
Kremer--Grest model \cite{AuhlEtal:2003}, and of a lattice model, the
bond-fluctuation model \cite{BaschnagelWittmerMeyer:NIC_Review2004}.
The basic idea of the algorithm is to find pairs of neighboring
chains which one can decompose into two halves and reconnect in a
way that preserves the monodispersity of the polymers.  Such a
connectivity-altering move drastically modifies the conformation of
the two chains and thus strongly reduces the slowing of the dynamics
due to large values of $\Np$.  However, if we attempt to repeat this
move over and over again on the melt configuration we started with,
a successful double-bridging event is likely to annihilate one of
its predecessors by performing the transition between two chains in
the reverse direction.  To avoid this inefficiency this nonlocal
chain updating should be complemented by a move which efficiently
mixes up the local structure of the melt.  At low $T$, efficient
relaxation of the liquid structure calls for a method which
alleviates the glassy slowing down in general.  Thus, any algorithm
achieving this aim in non-polymeric liquids should also accelerate
the equilibration of glassy polymer melts, provided that it can be
generalized to respect chain connectivity.  At present, no technique
has been established to fully solve this problem (see Ref.~\cite{BrumerReichman:JPC2004} for a topical review).
However, promising candidates appear to be `parallel tempering'
\cite{BunkerDunweg:PRE2000,YamamotoKob2000} (see however
\cite{DeMicheleSciortiono:PRE2002}), a recent MC approach proposed by Harmandaris and 
Theodorou combining scission and fusion moves \cite{Harmandaris2002a,Harmandaris2002b} or variants of
`Wang-Landau sampling' \cite{YanEtal:PRL2004,Binder2008}.

\section{Polymers at interfaces: some salient features}
\label{subsection:simulation-fene-polymer-in-film}
In order to demonstrate the capabilities of the simple model described above, we focus here on our recent simulation studies. Compared to the earlier Monte Carlo studies of the bond-fluctuation lattice model \cite{BaschnagelWittmerMeyer:NIC_Review2004}, reviewed in references~\cite{BinderBaschnagelPaul2003,MischlerEtal:ACIS2001}, our work deployed MD simulations to explore the features of a continuum model, spatially confined to a slab geometry \cite{VarnikEtal:PRE2002,VarnikEtal:EPJE2002,VarnikEtal:EPJE2003,VarnikBinder:JCP2002}.

\begin{figure}
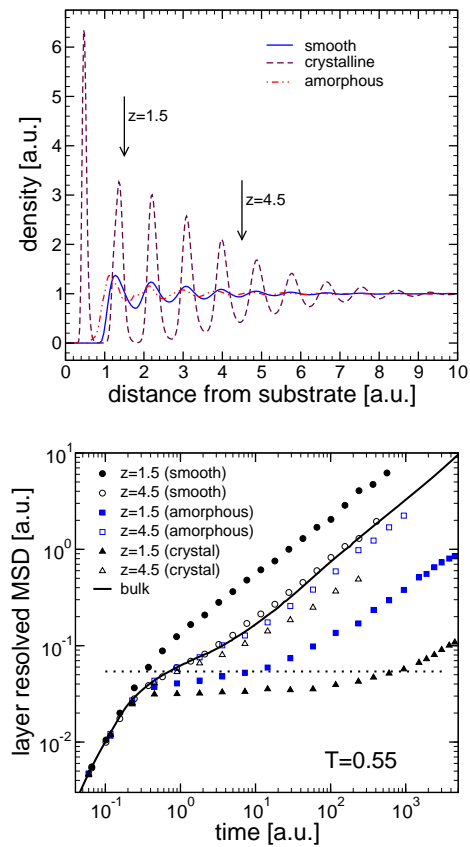

\hspace*{1mm}
\includegraphics[width=6cm]{density_profiles_p1_D20_T0.55_allWalls.eps}\vspace*{4mm}
\includegraphics[width=6cm]{msd_layer_resolved_allWalls.eps}
\caption[]{Molecular dynamics simulations of substrate effects on the structure (a) and dynamics (b) of a model polymer melt described by Eqs.~(\ref{eq:LJ12-6TS}) and (\ref{eq:FENE}). (a): Density profile (normalized to the liquid density at infinite distance) versus distance z from the substrate. (b): Local mean square displacements (MSD) versus time at two different distances, $z$, from a solid substrate \cite{BaVa2005}.}
\label{fig:rho+msd}
\end{figure}

As an example, \fref{fig:rho+msd} compiles results of molecular dynamics simulations of 
the polymer model of Eqs.\ (\ref{eq:LJ12-6TS}) and~(\ref{eq:FENE}) in the proximity of a solid substrate (at distances of a few monomer diameters), showing the strong heterogeneity induced by the substrate both in the static structure (exemplified by the local density) as well as in transport properties of the liquid. The figure also illustrates the influence of different solid structures such as a perfectly smooth substrate as compared to substrates with atomic scale corrugation (amorphous as well as crystalline).

As shown in \fref{fig:rho+msd}, the liquid density exhibits oscillations in the proximity of a substrate. These oscillations are of comparable magnitude both for an ideally flat and an amorphous corrugated substrate. Despite this similarity in the behavior of the local density, the effect on dynamics is of opposite nature. While a perfectly smooth substrate (no corrugation at all) leads to an acceleration of diffusive dynamics, the dynamics of monomers close to an amorphous substrate is slowed down. A qualitative understanding of this, at first glance unexpected, behavior can be obtained by invoking the concept of effective friction \cite{VarnikEtal:PRE2002,VarnikEtal:EPJE2002}.

\begin{figure}
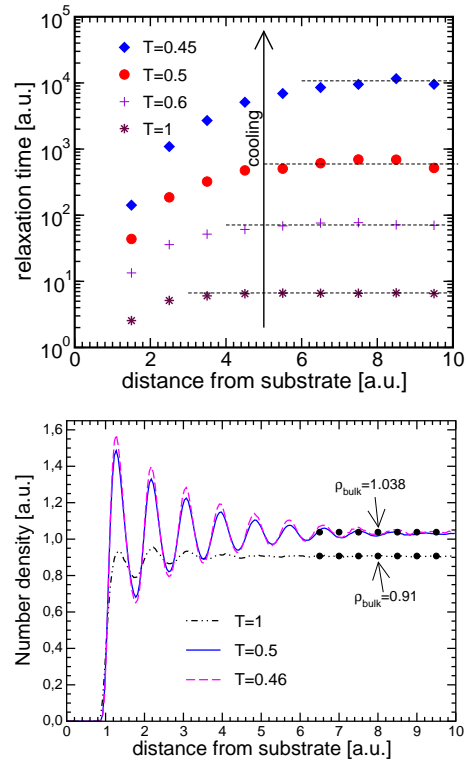

\vspace*{5mm}
\includegraphics[width=6cm]{fig18a.eps}\vspace*{3mm}
\includegraphics[width=6cm]{fig18b.eps}
\caption[]{Growth of both the strength and the range of substrate effects upon cooling.  (a): Relaxation time versus distance from the substrate for a model polymer melt described by Eqs.~(\ref{eq:LJ12-6TS}) and (\ref{eq:FENE}) at various temperatures (Lennard--Jones units). The horizontal dashed lines indicate the bulk values (expected at large distances from the substrate). (b): Monomer number density profile for the same range of temperatures \cite{VarnikEtal:PRE2002}.}
\label{fig:tau+rho}
\end{figure}

Figure \ref{fig:rho+msd} also underlines the fact that the spatial arrangement of substrate atoms may play a crucial role in the properties of the adjacent liquid. Strong layering of liquid particles has a dynamic counterpart manifest in a temporary arrest within the liquid layer closest to the substrate  (note the extended plateau in the mean square displacement).

Substrate effects both on the packing behavior as well as on dynamics can become quite dramatic when cooling the liquid towards the freezing temperature. This aspect is demonstrated in Fig.~\ref{fig:tau+rho} for the case of the polymer model of Eqs.~(\ref{eq:LJ12-6TS}) and~(\ref{eq:FENE}) close to a purely repulsive and perfectly smooth substrate (Eq.~(\ref{eq:LJ9-3}) with $\epsilon_\mathrm{w}=\epsilon$ and $f_\mathrm{w}=0$). The figure shows the structural relaxation time versus the distance from the substrate for various temperatures. While the effect of the substrate is rather weak and short ranged at high temperatures, the strength of the substrate effects grows significantly as the temperature is decreased. Furthermore, the spatial extension of the region affected by the substrate also becomes larger upon cooling. At the highest temperature shown, the substrate effects are visible only within a distance of 2--3 particle diameters. In contrast to this, the range of substrate effects exceeds 6 particle diameters at the lowest temperature investigated. The most remarkable effect, however, is on dynamic properties: At the lowest temperature investigated, the relaxation time decreases by roughly two orders of magnitude when approaching the substrate from infinity.

It is noteworthy that the above observations via computer simulations of a very simple model are qualitatively in line with experimental findings on real polymers \cite{Ellison2003,Ellison2005}. This indicates that, at least on a qualitative level, generic features such as geometric confinement as well as adhesive/repulsive nature of the substrate play a more important role than material specific details.

\begin{figure}
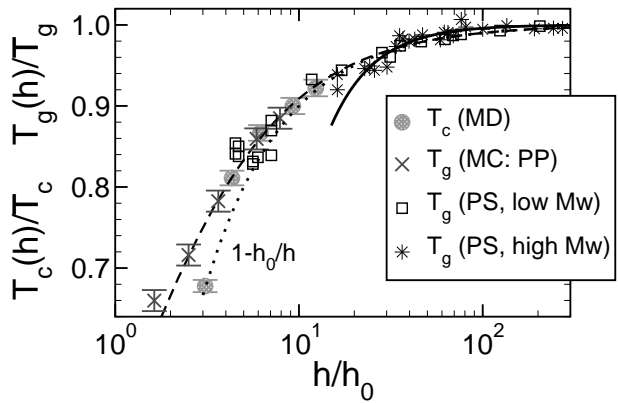
 \rsfig{0.45}{TcTg_vs_thickness_relative_values.eps}
\caption[]{Scaling plot of $\Tc(h)$ and $\Tg(h)$.
Here $h$ is the film thickness and $\Tc(h)$ is the so called critical temperature of the mode coupling theory determined in MD simulations of the polymer model specified via Eqs.~(\ref{eq:LJ12-6TS}) and (\ref{eq:FENE}) confined between two purely repulsive and perfectly smooth walls. It is compared to the glass transition temperatures $\Tg(h)$ of three studies: 
(i) Monte Carlo simulations of a lattice model for free-standing atactic polypropylene (PP) films \cite{XuMattice:Macro2003}
(ii) Experiments of supported atactic polystyrene (PS) films (spin cast
from toluene solution onto silicon wafers) \cite{HerminghausEtal:EPJE2001}
(iii) Experiments of supported, high-molecular weight PS films \cite{KeddieEtal:EPL1994,ForrestDalnoki:AdvColSci2001}. The solid line indicates $\Tg(h)/\Tg=1 - (h_0/{h})^\delta$ 
($h_0$ is a material dependent characteristic length) 
\cite{KimEtal:Langmuir2000,KimEtal:Langmuir2001}, the dashed line $ \Tg(h)/\Tg=1/(1+h_0/h)$ 
\cite{KeddieEtal:EPL1994} and the dotted line the approximation $1-h_0/h$, valid for small $h_0/h$.}
\label{fig:TcTg}
\end{figure}

The above discussed effects of substrate on the dynamics of structural relaxation have strong implications regarding the thermal and mechanical properties of the system.  Indeed, in the case of perfectly smooth and repulsive walls, the confined system behaves more liquid like compared to the same polymer system in the bulk (infinitely far from the substrate). Similarly, the presence of atomistically corrugated substrates with adhesive interactions increases the solid character of the system. For sufficiently small slab or film thickness ($h \lesssim 100$ nm), this behavior translates itself into a dependence of the glass transition temperature $\Tg$ (the temperature at which the polymer forms an amorphous solid) both on the type and thickness of the slab.
Figure \ref{fig:TcTg} illustrates this issue for the case of smooth non adhesive walls, where the expected reduction in $\Tg$ is observed in experiments accompanied by a similar reduction in $\Tc$ in simulations ($\Tc$ is the so called ideal glass transition temperature within the mode coupling theory \cite{GoetzeSjoegren1992_RPP,Goetze:MCTessentials,Goetze:JPCM1999}). It is noteworthy that, in the case of adhesive substrates, the expected opposite effect, i.e.\ an increase in the glass transition temperature upon confinement is indeed observed (see e.g.\ Torres and coworkers \cite{torresEtal:PRL2000} and references therein).

\begin{figure}
\vspace*{5mm}
\includegraphics[width=6cm]{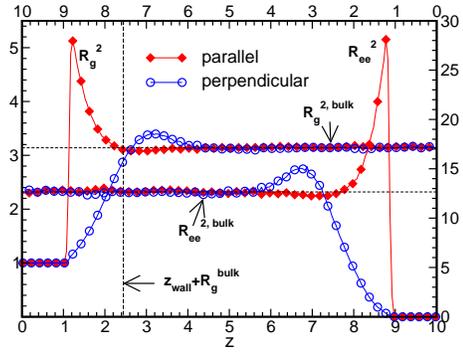}
\caption[]{Components of the radius of gyration and  the end-to-end distance in directions parallel ($\Rgpar$ and $\Reepar$) and perpendicular ($\Rgper$ and $\Reeper$) to the substrate versus the distance, $z$, from the substrate. $\Rgpar(z)$ (left ordinate) and $\Reepar(z)$ (right ordinate) behave qualitatively similar. They develop a maximum  close to the substrate and then converge towards a constant (bulk) value in the film center indicated by horizontal dashed lines. (Note that $\Rgpar$ and $\Rgper$ are shifted upwards by an amount of
unity in order to avoid crossing with the end-to-end curves)
\cite{VarnikEtal:EPJE2002}.}
\label{fig:fig4}
\end{figure}

Let us now turn to the question how polymer specific properties such as the conformation of a chain may change close to a substrate. For this purpose, we show in \fref{fig:fig4} that polymer conformation is significantly stretched along the parallel direction in the proximity of the substrate. For this purpose, we compare the parallel and perpendicular components of the radius of gyration, $\Rgpar$, $\Rgper$ as well as the same components of the end-to-end distance, $\Reepar$, $\Reeper$ in a slab geometry.

The components are plotted versus the distance, $z$, from the substrate, where $z$ denotes the position of chain's center of mass.  So, $\Rgpar(z)$, for instance, is the radius of gyration parallel to the substrate, which is averaged over all chains whose centers of mass are located at $z$. The figure shows that both the radius of gyration and the end-to-end distance agree with the bulk value if $z > z_{\rm w} \!+\! 2 \Rgbulk$. Here, $z_{\rm w} \approx 1$ is the wall position, i.e.\ the smallest distance between a monomer and the substrate. As the chain's center of mass approaches the substrate, $\Rgpar$ and $\Reepar$ first develop a shallow minimum and then increase to about twice the bulk value followed by a sharp decrease to zero in the very vicinity of the substrate where practically no chain is present. On the other hand, the perpendicular components, $\Rgper$ and $\Reeper$, first pass through a maximum before decreasing to almost 0 at the substrate. This behavior has been observed in several other simulations  (see \cite{BaschnagelBinderMilchev2000} and references therein), also for larger chain length than studied here \cite{BitsanisHadziioannou1990,WangBinder1991}.

\section{Conclusion}
In this paper, we provide a brief survey of modeling and simulation studies of dense systems of flexible linear polymers close to solid substrates. The challenging character of polymer science relies upon the fact that the smallest and largest length scales present in a polymer system may span many orders of magnitude, usually $\sim 1 \AA$ (size of an atom) up to hundreds of nanometers (end-to-end distance). This broad range of length scales brings about a correspondingly wide time window. An adequate study of polymer systems thus necessarily involves the use of simplifying concepts allowing one to focus on essential features. It is, therefore, not surprising that coarse graining procedures are common practice in polymer science. We therefore address in this article some of the basic ideas regarding the development of coarse grained models for a computational study of polymer systems. Along this route, we also address Monte Carlo methods as compared to molecular dynamics simulations.

Some salient properties of polymers close to a solid substrate are also presented. In particular, we show how the presence of a solid substrate may affect both the static and dynamic properties of a polymer melt. This is exemplified by a significant slowing down of the local diffusion dynamics  --and the closely related dynamics of structural relaxation-- in the proximity of attractive substrates. Similarly, a generic enhancement of diffusion (resulting in a reduction of the glass transition temperature in a slab geometry) is observed close to perfectly smooth non adhesive surfaces. These findings are evidenced both experimentally and by computer simulations. Polymer specific features, on the other hand, reflect themselves, e.g.\ in a change of conformational degrees of freedom from fully isotropic in the bulk to an elongated (in direction parallel to the substrate) state in the very vicinity of the substrate.

We gratefully thank J. Baschnagel, Wolfgang Paul and Marcus M\"uller for useful discussions and providing us with results of their recent work. FV acknowledges the support from the industrial sponsors of the ICAMS, the state of North Rhine-Westphalia and European Union.






\end{document}